\documentclass[12pt]{article}
\usepackage{Sc3conf}
\usepackage{amsfonts}
\usepackage{epsf}

\begin{document}
\def\csch{\mathop{\rm csch}\nolimits}
\def\lss{\mathop{<}\limits_{\sim}}
\def\ds{\displaystyle}
\def\ts{\textstyle}
\def\ss{\scriptstyle}
\def\d{\partial}
%%%%%%%%%%%%%%%%%%%%%%%%%%%%%%%%%%%%%%%%%%%%%%%%%%%%%%%%%%%%%%%%%%%%%%%
%%%  The TITLE of the paper

\title{The mass-shift method and the self-action of classical charge}

%%%%%%%%%%%%%%%%%%%%%%%%%%%%%%%%%%%%%%%%%%%%%%%%%%%%%%%%%%%%%%%%%%%%%%%
%%%  The AUTHORS.  If some of the authors belong to different
%%%  institutions, the addresses of those institutions (which appear
%%%  below) are referred from here by commands \1ad, \2ad, etc., or
%%%  the command \adref{..}

\authors{S.L.~Lebedev,\adref{1} }

%%%%%%%%%%%%%%%%%%%%%%%%%%%%%%%%%%%%%%%%%%%%%%%%%%%%%%%%%%%%%%%%%%%%%%%
%%%  The ADDRESSES.  If there is more than one address, they are
%%%  prepended after the \nextaddress command

\addresses{\1ad Chuvash state pedagogical university, 428000 Cheboksary}

%%%%%%%%%%%%%%%%%%%%%%%%%%%%%%%%%%%%%%%%%%%%%%%%%%%%%%%%%%%%%%%%%%%%%%%
%%% The mandatory command \maketitle actually typesets the title,
%%% author names and affiliations using the definitions above.

\maketitle

%%%%%%%%%%%%%%%%%%%%%%%%%%%%%%%%%%%%%%%%%%%%%%%%%%%%%%%%%%%%%%%%%%%%%%%
%%%  The abstract shouldn't contain more than 20 lines when printed.

\begin{abstract}
The complex non-local action functional is used in classical electrodynamics
to describe the back-reaction effects for the charge moving in the constant
homogeneous electromagnetic field. We apply the mass-shift method to obtain
the higher order radiation effects in the non-relativistic cyclotron motion
and generalize the method to the case of Bargmann-Michel-Telegdi particle.
\end{abstract}

%%%%%%%%%%%%%%%%%%%%%%%%%%%%%%%%%%%%%%%%%%%%%%%%%%%%%%%%%%%%%%%%%%%%%%%
%%%  The body of the document begins here.

\section{Introduction}
In 1978, considering the mass operator of the electron in the constant
homogeneous EM field, V.~Ritus \cite{R78} found the field-dependent correction
to the mass of the electron, which in the weak-field limit $\beta \ll 1$ ($\beta
\equiv {e\hbar \varepsilon}/{m^2 c^3}$, $\varepsilon$ -
the electric field) takes the form \footnote{With some obvious exceptions
we use the system of units where
$c=1$, $\hbar=1$. Fine structure constant $\alpha =e^2 / 4\pi \hbar c$.}:
$$
\Delta m=\frac{\alpha m}{2\pi}\left[-\pi\beta - i\beta
\left(2\ln \frac{2\beta m}{\gamma_E \mu_{ph}} -1\right)+\dots\right] .
\eqno{(1)}
$$
Here $\ln\gamma_{E}=0.577\dots$ and we have omitted the quantum corrections
of orders $\beta^2 \ln\beta$ and $\beta^2$. The leading terms in the real
and imaginary parts of the mass shift (MS), as it can be deduced from (1),
are purely classical \footnote{The IR parameter $\mu_{ph}$ ("photon mass")
should be interpreted as a minimal wave number $k_{\perp
min}=\mu_{ph}c/\hbar$ of radiating quanta \cite{R8182}.}. Subsequently, Ritus
\cite{R8182} suggested a method of calculation of the MS (1) which relies,
entirely, on a classical quantity
$$
\Delta W=\left. \frac12 e^2\,\int d\tau\int
d\tau^{'}\dot x_{\alpha}(\tau)\dot x_{\alpha}(\tau^{'})
\Delta_c (x-x^{'};\mu_{ph})\right |_{0}^{F}  \, .
\eqno{(2)}
$$
$\Delta W$ is the change in the self-action of the charge caused by the external
field. The overdots in (2) denote the derivatives of the world line
$x_{\alpha}(\tau)$ w.r.t. the proper time.

In the constant homogeneous field the self-action (2) determines the MS
according to
$$
\Delta W=-\Delta m T \, ,
\eqno{(3)}
$$
where the proper time interval $T$ of the motion (in external field) is much
greater than the formation time of $\Delta m$.

The Green's function $\Delta_c$ in (2) is the causal one and, consequently,
$\Delta W$ is in general the complex quantity. The next important point is
the presence of the subtraction $\left. \right|_0^F$ which manifests the
fact that all self-action effects taken at zero external field should be
accounted for by the definition of the observable mass of the charge. This
means that no UV divergences should appear in (2).

To interpret $\Im\Delta W$ in the context of classical electrodynamics
\cite{R8182} (CED) one ought to remind the connection between the latter and the
spectral function of radiation,
$$
d{\mathcal E}_{\bf k}=|j_{\mu}({\bf k})|^2\,\,\frac{d^3{\bf k}}{16\pi^3}\,\,\,
\,\,
,\frac{1}{\hbar}\Im\Delta W=\int\,\frac{d{\mathcal E}_{\bf k}}{\hbar \omega}\,,
\eqno{(4)}
$$
produced by the classical source (see e.g. \cite{IZ})
$$
j_{\mu}(x)=e\int \,d\tau \dot x_{\mu}(\tau)\delta^{(4)}(x-x(\tau))\, .
\eqno{(5)}
$$

The quantity $\exp\left(\frac{i}{\hbar}\Delta W\right)$ appears in QED as
well in the form of an amplitude of the preservation of the photon vacuum
in the presence of the classical source (5), so that the imaginary part of
$\Delta W$ determines the corresponding probability \cite{IZ}:
$$
\exp(-2\Im\Delta W)=|<0^+|0^->_j|^2\,\,.
\eqno{(6)}
$$
This probability is less than unity if the particle radiates.

The consideration of the radiation reaction effects through the non-local
reparametrization invariant functional (2) could bring the information which
is supplementary to one obtained from Abraham-Lorentz-Dirac (ALD) equation.
Let us list some known applications of the formula (2) (this applications,
certainly, are not restricted with the case of constant electric field):
%\begin{itemize}
%\item

\noindent {--the self-energy of the charged particle rest in a proximity to the
charged black hole. MS determines a repulsive force acting on this particle
\cite{FZ};}
%\item

\noindent{--the real part of $\Delta m$ in (1) determines (at $\beta \ll 1$) the
correction to the classical action in the exponential of barrier factor.
This results \cite{R84} in 2-loop radiative correction to Schwinger pair
creation rate;}

%\item
\noindent{--the classical radiative corrections to the cyclotron motion of the
electron near a boundary \cite{LS94}. Those were in the focus of attention
in connection with the apparatus-dependent effects in $g-2$ experiments of
Washington group \cite{BG}.}
%\end{itemize}
It should be stressed that in the second example the result could not be
deduced from ALD equation in view of infrared origin of the real part of
the MS \cite{R8182}.

\section{Uniformly accelerated charge}
 Here we consider the MS of the uniformly accelerated (UA) charge radiating
vector (that is EM) or scalar field (electric or scalar charge for brevity)
and moving in the space-time of arbitrary dimensions $D$. For $D=4$ this
type of motion is distinguished by zero radiation reaction force, so that
the world line $x(\tau)$ of UA charge is an exact solution of the ALD equation.
At that time one finds (see \cite{LY}) that the value $D=4$ is in a sense
singular for the existence of the ALD-type equation. We find the corresponding
pecularities for the MS as well.

To begin with we remind one remarkable similarity ($D=4$) between the mass
shifts of electric charge considering at two different situations. a)The
self-field is massive (see (1)-(3)),
$$
\Delta m_{el}=\frac{\alpha
w_0}{2\pi}\,S_{el}(\Lambda),\,\,
\Lambda=\mu_{ph}/w_0 \equiv k_{\perp min}/w_0
\eqno{(7)}
$$
($\Re S_{el}$ and $\Im S_{el}$ are expressed \cite{R8182} through the bilinear
combinations of Bessel functions, $w_0$ is an acceleration in the rest frame
of the charge and parameter $\beta$ from (1) is $\beta=w_0/m$); b) the
self-field is massless but the charge is moving parallel to and a distance
$L$ apart from the perfect mirror,
$$
\Delta m_{el}=\frac{\alpha
w_0}{2\pi}\,V_{el}(\bar\Lambda),\,\,\,\,
\bar\Lambda=(\gamma_ELw_0)^{-1}
\eqno{(8)}
$$
(Euler's constant is used here to fix the correspondence between $S$ and
$V$ in the IR limit). The function $V_{el}$ is an elementary one \cite{LS89}:
$$
V_{el}=\pi\left(-\coth{\theta}+\frac12 \csch{\frac{\theta}{2}}\right)
+i(1-\theta\coth{\theta}),\,\, \cosh{\theta}=1+2(\bar\Lambda \gamma_E)^{-2}.
\eqno{(9)}
$$
The correspondence mentioned above,
$$
\Lambda \leftrightarrow \bar\Lambda\simeq (Lw_0)^{-1},
\eqno{(10)}
$$
is fulfilled (as it can be seen from the Fig.1) in the wide interval
$$
0<\Lambda,\bar\Lambda \lss 10 ,
\eqno{(11)}
$$
despite that only in the IR region $\Lambda,\bar\Lambda \ll 1$ the asymptotics
of $S_{el}$ and $V_{el}$ are expected to be in common (for real parts, $\Re
S_{el}$ and $\Re V_{el}$, this similarity extends over the whole positive
axis).

For the scalar UA source we have the same correspondence (10) between
$S_{sc}(\Lambda)$ and $V_{sc}(\bar\Lambda)$ (Fig.1b). The IR limit of those
functions is distinct from (1) \cite{R8182,FZ,LS89}:
$$
\Delta m_{sc}=-i\frac{\alpha w_0}{2\pi}+\dots \,(\Lambda,\bar\Lambda\ll 1).
\eqno{(12)}
$$
Now, in addition to abovementioned motivation it is interesting to consider
the MS $\Delta m_{el}$ and $\Delta m_{sc}$ as a functions of another IR
regulator (the latter being now $D-4$).
\begin{figure}
%  \begin{center}
%     \includegraphics[bb = 0 0 2cm 1.5cm,
%       height=3cm, keepaspectratio]{dummy1.eps}
    \epsfxsize =7cm \epsfbox{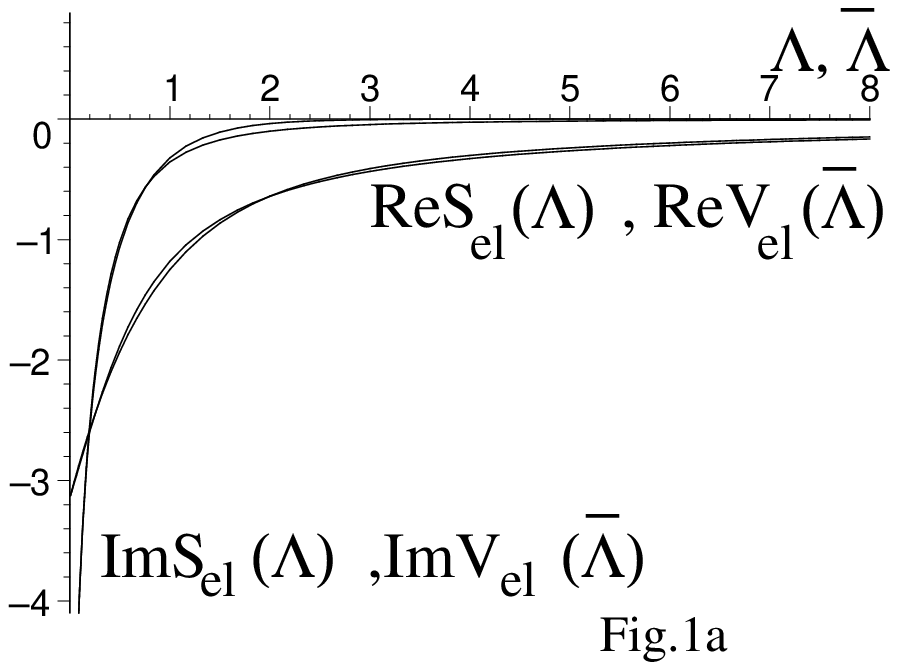}
\epsfxsize =7cm \epsfbox{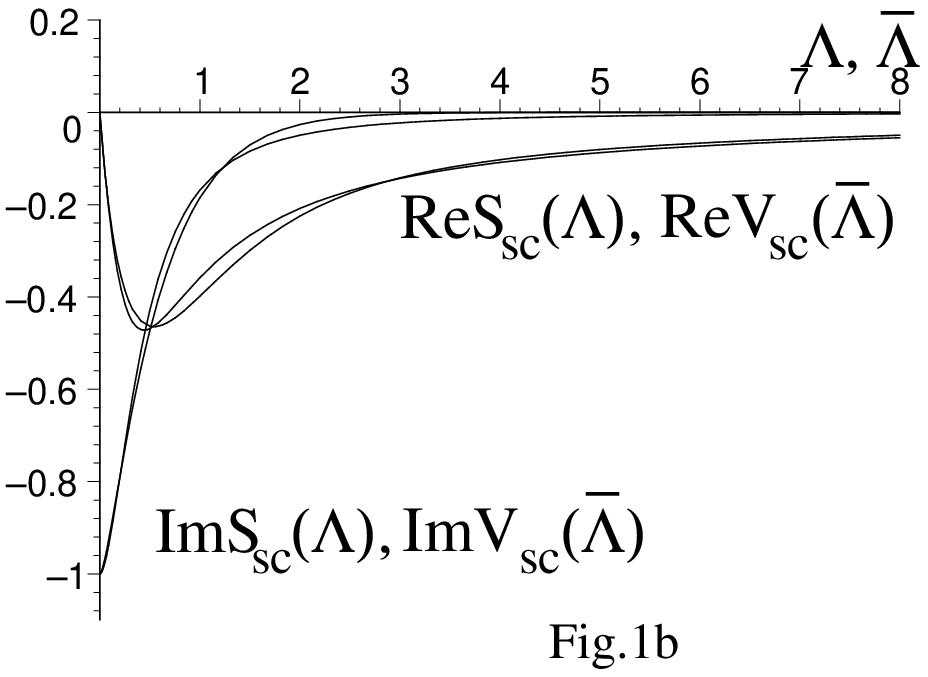}
%    \caption{Just a figure, not necessarily requiring a
%      caption}\label{fig:1}
%\vspace{5mm}
 % \end{center}
\end{figure}

We need usual dimensional generalization of $\Delta_c$ in the self-action
(2) for $D\neq 4$ and introduce mass scale $\mu$. Now the source (5) has
the charge ($e$--dimensionless constant)
$$
e'=e\mu^{\frac{4-D}{2}}
\eqno{(13)}
$$
and $D$-dimensional $\delta$-function is implied there. The simple calculations
along the lines of the work \cite{R8182} result in the expression:
$$
 {\Delta m_{el}=-\frac{e^2w_0}{2(2\pi)^\frac D2}
\left(\frac{w_0}{\mu}\right)^{D-4}\exp{\textstyle (\frac{-i\pi D}{4})}
\int\limits_0^{\infty}dz\,z^{\frac{D-4}{2}}\,e^{iz}K_1(iz)\,} .
\eqno{(14)}
$$
The corresponding formula for the $\Delta m_{sc}$ could be obtained from
(14) by the substitution $(-K_0(iz))$ in place of $K_1(iz))$ ($K_{\nu}$'s
are the McDonald functions). The integral in (14) is transformed
with the help of 2.16.6(5) in \cite{PBM2}, that gives the final expressions:
$$
{\Delta m_{el}=\frac{e^2w_0}{(4\pi)^\frac D2}
\left(\frac{w_0}{\mu}\right)^{D-4}\frac{i}{\sqrt\pi}
\exp{\textstyle (\frac{-i\pi D}{2})}\textstyle{
\Gamma\left(\frac{D}{2}\right)\Gamma\left(\frac{D-4}{2}\right)
\Gamma\left(\frac{3-D}{2}\right)\,} ,}
\eqno{(15)}
$$
$$
\Delta m_{sc}=\frac{e^2w_0}{(4\pi)^\frac D2}
\left(\frac{w_0}{\mu}\right)^{D-4}\frac{i}{\sqrt\pi}
\exp{\textstyle (\frac{-i\pi D}{2})}\textstyle{
\Gamma\left(\frac{D-2}{2}\right)\Gamma\left(\frac{D-2}{2}\right)
\Gamma\left(\frac{3-D}{2}\right)\,} .
\eqno{(16)}
$$
At $2\epsilon\equiv D-4 \rightarrow +0$ one finds
$$
{\Delta m_{el}=\frac{\alpha w_0}{2\pi}\left[-\pi -
i\left(\frac{1}{\epsilon}+
\ln{\frac{w_0^2}{\pi\mu^2\gamma_E}} -1\right)\right]},
\eqno{(17)}
$$
and $\Delta m_{sc}$ in the same limit is precisely (12). With the
correspondence $\mu_{ph}^2=4\pi\mu^2\gamma_E^{-1}\exp{(-1/\epsilon)}$
expression (17) can be easily reconciled with (1).

The following properties are characteristic of the MS (15) and (16): i)the
real part of $\Delta m_{el}$ in (17) appears in tandem with the IR singularity
$\epsilon^{-1}$ present in (15), precisely as it were for the MS (7) and
(8); ii) the real parts $\Re\Delta m_{el}$ and $\Re\Delta m_{sc}$ do not appear
for even dimensions $D\ge 6$; iii) for odd $D\ge 3$ the real parts
$\Re\Delta m_{el}$ and $\Re\Delta m_{sc}$ are divergent, while imaginary
ones are finite. This should be juxtaposed with the absence of the Huygence
principle in odd-dimensional space-times.

\section{Cyclotron motion}
For the non-relativistic cyclotron motion in magnetic field ($D=4$ below)
one knows an exact solution of the ALD equation,
$$
{\frac{d\xi}{dt}-\frac1b \frac{d^2\xi}{dt^2}=-i\omega_c\xi},
\eqno{(18)}
$$
where the following notations have been introduced:
$$
{\xi=v_x+iv_y,\,\,\,\omega_c=\frac{eH}{m},\,\,\, \frac1b =\frac23
\frac{e^2}{4\pi m}}.
\eqno{(19)}
$$
We shall discuss only the plane motion of the charge. The non-relativistic
cyclotron frequency $\omega_c$ describes unperturbed motion without radiation
 damping, present as the second term in the l.h.s. of (18). Physically
acceptable (and exact) solution of (18) was found in \cite{GPl} and looks like
$$
{\xi=A\exp(-i\Omega t)}
\eqno{(20)}
$$
$$
{\Re\Omega=\frac{b}{2}\left(-\frac12+\frac12
\sqrt{1+16\frac{\omega_c^2}{b^2}}\right)^{1/2}=
\omega_c\left(1-2\frac{\omega_c^2}{b^2}+\dots\right)},
\eqno{(21)}
$$
$$
{\Im\Omega=\frac{b}{2}\left[1-\left(\frac12+\frac12
\sqrt{1+16\frac{\omega_c^2}{b^2}}\right)^{1/2}\right]=
-\frac{\omega_c^2}{b}+\dots}\,\,.
\eqno{(22)}
$$
Note, that real and imaginary parts of $\Omega$ are represented here as an
infinite series in radiative constant $(\omega_c/b)^2$ and that $\Im\Omega$
is negative (this property corresponds to the dissipation of the energy and
to the shrinkage of the cyclotron orbit).

MS for the plane cyclotron motion was found in the aforementioned Ritus's
work \cite{R8182} and is purely imaginary quantity \footnote{MS (23) is not
an exact result in the sense that $\Delta W$ (2) is computed for the
unperturbed cyclotron world line.}:
$$
{\Delta m=-i\frac{\alpha\omega_c}{2\pi}\int\limits_0^\infty dx\,
\left(\frac{1-v_{\perp}^2 \cos{2x}}{x^2-v_{\perp}^2\sin^2{x}}-\frac1{x^2}
\right)\equiv-i\frac{\alpha\omega_c}{2\pi}I(v_{\perp}^2)}\,.
\eqno{(23)}
$$
Now we shall exploit the dependence of $\Delta m$ on the integral
$v_{\perp}^2\equiv v_x^2+v_y^2$ of the unperturbed motion. The first order
correction to Lagrange function in laboratory system, $\Delta L=-\frac{\Delta
m}{\gamma_{\perp}}$, in non relativistic regime takes the form:
$$
\Delta L=\frac12 \delta m^{(1)}v_{\perp}^2 \,\, ,
\eqno{(24)}
$$
where the first-order (in $e^2$) radiative correction to the mass,
$$
\delta m^{(1)}=\frac{2i\alpha}{3m}eH
\eqno{(25)}
$$
($\delta m^{(1)}$ should not be confused with $\Delta m$: having different
physical meaning, they differ in sign). We postulate that the correction up
to the second order must be determined by the same principle, but with
$m+\delta m^{(1)}$ substituted in place of m in (25). Application of this
procedure {\it ad infinitum} gives us the equation for the complete radiative
mass-correction
$$
\delta m=\frac{2i\alpha eH}{3(m+\delta m)} \,.
\eqno{(26)}
$$
It could be easily checked that the "new" cyclotron frequency
$eH/(m+\delta m)$ is just the $\Omega$ from (21), (22):
$$
\frac{\delta m}{m}=-\frac12 + \sqrt{\frac14+\frac{i\omega_c}{b}}\,\,\,,
\Omega=\frac{eH}{m+\delta m}\, .
\eqno{(27)}
$$
Qualitatively, this coincidence is not surprising: radiation effects, when
properly accounted, must entail in dynamics. Really unexpected is quantitative
correspondence between formulas (27) and (21), (22). So, one should ask about
the foundation of such "empirical" perturbative approach. As it seems now
the main points are the following:
i) the structure of the radiative correction (24) to Lagrange function (or
energy) is just the one of the "bare" Lagrangian. $\delta m^{(1)}$
does not contain dynamical variables, so that the relation (25) is independent
of the concrete conditions of the cyclotron motion; ii) the constancy of the
field which makes going from non-local $\Delta W$ to the MS possible;
iii) two dimensional character of the unperturbed motion  is not destroyed by
radiation; the latter does give the better conditions for the applicability
of the non-relativistic approximation. Could one hope to use this
renormalisation method as a general tool for solving ALD equation? It does
not seem so because there is no universal dimensionless parameter in CED
which would govern the radiation effects. QED does have such a parameter
which is the fine structure constant.

\section{The probe of radiation reaction effects for spinning particles}
Several reasons make this topic interesting: i) considerable recent attention
to the pseudoclassical models which stems from the close relations between
those models and string theory; ii) with rare exception, the problem
of back-reaction effects have not been considering in those models; iii)
it is desirable to endow the spinning particles models with a quasiclassical
status -in the same sense as one uses the BMT equation (which does not include
back reaction). For example, there is still a need in adequate quasiclassical
interpretation of the radiation polarization phenomenon \cite{Ter}.

The self field of the charged particle possessing the magnetic moment $\mu$
($=\frac{g}{2} \mu_B,\,\,\mu_B\equiv e\hbar/2mc$) satisfies the equation
$$
-\d^2 A_{\beta}(x)=j_{\beta}(x)+\d_{\gamma}M_{\beta\gamma}(x)\,\, ,
\eqno{(28)}
$$
where $j_{\beta}(x)$  is the source (5), and
$$
M_{\beta\gamma}(x)=\int d\tau \,\mu_{\beta\gamma}(\tau)
\delta^{(4)}(x-x(\tau))
\eqno{(29)}
$$
is the polarization density. The dependence of
$$
\mu_{\beta\gamma}=i\mu \varepsilon_{\beta\gamma\varepsilon\delta}
\dot x_{\varepsilon}S_{\delta}
\eqno{(30)}
$$
on $\tau$ is determined from the Lorentz and BMT equations \cite{IZ}:
$$
\ddot x_{\alpha}=2\mu_BF_{\alpha\beta}\dot x_{\beta}\,\,,
\dot S_{\alpha}=2\mu_B F_{\alpha\beta}S_{\beta}
\eqno{(31)}
$$
(we put g=2 for simplicity). The self-action obtained from (2) by substitution
$j_{\beta}\rightarrow j_{\beta}+\d_{\gamma}M_{\beta\gamma}$, can be decomposed
into the following terms:
$$
\Delta W=\Delta W_{or}+\Delta W_{s-o}+\Delta W_{s-s}\, ,
\eqno{(32)}
$$
where "orbit" part is precisely $\Delta W$ in (2), and "spin-orbit" and
"spin-spin" terms are:
$$
\Delta W_{s-o}=\left.-e\int d\tau \int d\tau'\dot x_{\beta}(\tau)
\,\mu_{\beta\alpha}(\tau ')\d_{\alpha}^{'}\Delta_c(x-x')\right |^F_0 \,,
\eqno{(33)}
$$
$$
\Delta W_{s-s}=\left.\frac12 \int d\tau \int d\tau '\mu_{\alpha\beta}\,
\mu_{\alpha\gamma}^{'}\d_{\beta}\d_{\gamma}^{'}\Delta_c(x-x')\right |^F_0 \,.
\eqno{(34)}
$$

For the constant homogeneous magnetic field the MS is expressed through the
geometrical invariants of the world line, the latter being curvature $k$
and the first torsion $\tau_1$ (the second one is equal to zero for the plane
motion):
$$
k=2\mu_B Hv_{\perp}\gamma_{\perp}\,\,\,\,\,\,,\tau_1=2\mu_B H\gamma_{\perp}\,.
\eqno{(35)}
$$
Without going into details, we give the final expression for the
$\Delta W_{s-o}$ term in (33) regarding the latter as a major contribution
between two terms (33) and (34) in the decomposition (32):
$$
\Delta m_{s-o}=-i\frac{e\mu}{2\pi^2}\zeta \omega^2_c f(v_{\perp}).
\eqno{(36)}
$$
Here $\zeta$ is $z$-component of spin (in $\hbar /2$ units), and the
formfactor $f(v_{\perp})$ includes relativistic retardation effects.
It is obtained from (33) (where IR regulator $\mu_{ph}$ could be omitted):
$$
f(v_{\perp})=2v^2_{\perp}\gamma_{\perp}^{-1}\int\limits_0^\infty dx\,
\frac{4\sin^2{(x/2)}-x\sin{x}}{[x^2-4v_{\perp}^2\sin^2{(x/2)}]^2}\, .
\eqno{(37)}
$$
%The plot of this function is presented on the Fig.3.
Following asymptotic expressions are obtainable from the representation (37):
$$
f(v_{\perp})=\left\{\begin{array}{cc}\frac{\pi}{6}v_{\perp}^2
\,\,\,,v_{\perp}\ll 1, \\  \\
\frac{\pi}{4\sqrt{3}}\gamma_{\perp}^2
\,\,\,\, , \gamma_{\perp}\gg 1 \, .
\end{array} \right.
\eqno{(38)}
$$
Comparing it with the ultrarelativistic behaviour of function $I(v_{\perp}^2)$
introduced in (23), $I(v_{\perp}^2)=\frac{5\pi}{2\sqrt{3}}\gamma_{\perp}$,
we can conclude about the notable growth of the spin effects
relative to orbit ones in the ultrarelativistic region $\gamma_{\perp}\gg 1$:
$$
\frac{\Delta m_{s-o}}{\Delta m_{or}}\simeq\frac15 \frac{H}{H_c}\,
\gamma_{\perp}\,.
\eqno{(39)}
$$
Here $H_c=m^2c^3/e\hbar$ and we put $\zeta \sim 1$ (see (23) and (36)).

\noindent
\textbf{Acknowledgments.}  {The author is grateful to A.I.~Nikishov and
V.I.~Ritus for useful discussions and to RFBR for partial financial support
(grant 00-15-96566)}.

%%%%%%%%%%%%%%%%%%%%%%%%%%%%%%%%%%%%%%%%%%%%%%%%%%%%%%%%%%%%%%%%%%%%%%%
%%%  The Bibliography is set using standard LaTeX macros

\end{document}